\documentclass[%
reprint,
superscriptaddress,
nofootinbib, %%%%
nobibnotes, %%%%
amsmath,amssymb,
aps,
prl,
]{revtex4-2}

\usepackage{graphicx}% Include figure files
\usepackage{dcolumn}% Align table columns on decimal point
\usepackage{bm}% bold math
\usepackage{xcolor}
\usepackage{braket}
\usepackage{siunitx}
\usepackage{natbib}
\usepackage{url}

\newcommand{\kkb}{$\text{K} \cdot k_B$}

\newcommand{\Be}{$^9$Be$^+$}
\newcommand{\z}{\text{z}}
\newcommand{\kb}{k_\text{B} }
\newcommand{\subf}[2]{%
	\setlength{\tabcolsep}{0pt} % space between e.g. a) and figure as small as possible
	{\small\begin{tabular}[t]{cc}
			{\fontfamily{phv}\selectfont 
				#2} & \raisebox{-0.95\height}{ #1} \\
		\end{tabular}
	}%
	\setlength{\tabcolsep}{0pt} % setting back to default value
}

\graphicspath{{./figures/}}

\begin{document}
	
	\preprint{APS/123-QED}
	
	\title{Image-current mediated sympathetic laser cooling of a single proton in a Penning trap down to 170\,mK axial temperature}% Force line breaks with \\

\newcommand{\mpik}{Max-Planck-Institut für Kernphysik, Saupfercheckweg 1, D-69117 Heidelberg, Germany}
\newcommand{\riken}{RIKEN, Fundamental Symmetries Laboratory, 2-1 Hirosawa, Wako, Saitama 351-0198, Japan}
\newcommand{\utexas}{Department of Physics, The University of Texas at Austin, Austin, Texas 78712, USA}
\newcommand{\jgu}{Institut für Physik, Johannes Gutenberg-Universität, Staudingerweg 7, D-55128 Mainz, Germany}
\newcommand{\uhanov}{Institut f{\"u}r Quantenoptik, Leibniz Universität Hannover, D-30167 Hannover, Germany}
\newcommand{\ptb}{Physikalisch-Technische Bundesanstalt, D-38116 Braunschweig, Germany}
\newcommand{\cern}{CERN, 1211 Geneva, Switzerland}
\newcommand{\utokyo}{Graduate School of Arts and Sciences, University of Tokyo, Tokyo 153-8902, Japan}
\newcommand{\gsi}{GSI Helmholtzzentrum für Schwerionenforschung GmbH, D-64291 Darmstadt, Germany}
\newcommand{\helmholtzMz}{Helmholtz-Institut Mainz, D-55099 Mainz, Germany}
\newcommand{\ethzurich}{Eidgen{\"o}ssische Technische Hochschule Z{\"u}rich, John-von-Neumann-Weg 9,
	8093 Zürich, Switzerland}
\newcommand{\hhu}{Heinrich Heine University, D{\"u}sseldorf, Universit{\"a}tsstrasse 1, D-40225 D{\"u}sseldorf, Germany}
\newcommand{\imperial}{Centre for Cold Matter,
	Blackett Laboratory, Imperial College London, Prince Consort Road, London SW7 2AZ, UK}

\author{C. Will}
\affiliation{\mpik}

\author{M. Wiesinger}
\affiliation{\mpik}

\author{P. Micke}
\altaffiliation{Present address: \textit{ Helmholtz Institute Jena, GSI Helmholtz Centre for Heavy Ion Research, Planckstraße 1, 64291 Darmstadt, Germany} }
\affiliation{\mpik }
\affiliation{\cern }

\author{H. Yildiz}
\affiliation{\jgu }

\author{T. Driscoll}
\altaffiliation{Present address: \textit{The University of Oregon, Eugene, Oregon 97403, USA} }
\affiliation{\mpik}
\affiliation{\utexas}

\author{S. Kommu}
\affiliation{\jgu }

\author{F. Abbass}
\affiliation{\jgu}

\author{B. P. Arndt}
\affiliation{\mpik}
\affiliation{\riken}
\affiliation{\gsi}

\author{B. B. Bauer}
\affiliation{\jgu}

\author{S. Erlewein}
\affiliation{\cern}
\affiliation{\riken}

\author{M. Fleck}
\affiliation{\riken}
\affiliation{\utokyo}

\author{J. I. Jäger}
\affiliation{\mpik}
\affiliation{\cern}
\affiliation{\riken}

\author{B. M. Latacz}
\affiliation{\cern}
\affiliation{\riken}

\author{A. Mooser}
\affiliation{\mpik}

\author{D. Schweitzer}
\affiliation{\jgu}

\author{G. Umbrazunas}
\affiliation{\riken}
\affiliation{\ethzurich}

\author{E. Wursten}
%	\affiliation{\cern}
\affiliation{\riken}

\author{K. Blaum}
\affiliation{\mpik}

\author{J. A. Devlin}
\affiliation{\imperial}

%	\author{Y. Matsuda}
%	\affiliation{\utokyo}

\author{C. Ospelkaus}
\affiliation{\uhanov}
\affiliation{\ptb}

\author{W. Quint}
\affiliation{\gsi}

\author{A. Soter}
\affiliation{\ethzurich}

\author{J. Walz}
\affiliation{\jgu}
\affiliation{\helmholtzMz}

\author{C. Smorra}
\affiliation{\jgu}

\author{S. Ulmer}
\affiliation{\riken }
\affiliation{\hhu}

\collaboration{BASE Collaboration}

\date{\today}% It is always \today, today,

\begin{abstract}
	We demonstrate a new temperature record for image-current mediated sympathetic cooling of a single proton in a cryogenic Penning trap by laser-cooled \Be. An axial mode temperature of 170\,mK is reached, which is a 15-fold improvement compared to the previous best value.
	Our cooling technique is applicable to any charged particle, so that the measurements presented here constitute a milestone towards the next generation of high-precision Penning-trap measurements with exotic particles.
\end{abstract}
\maketitle

\begin{figure}
	\centering
	\begin{tabular}{ll}
		\rule{0pt}{-1.0ex}%
		\includegraphics[width=8.6cm]{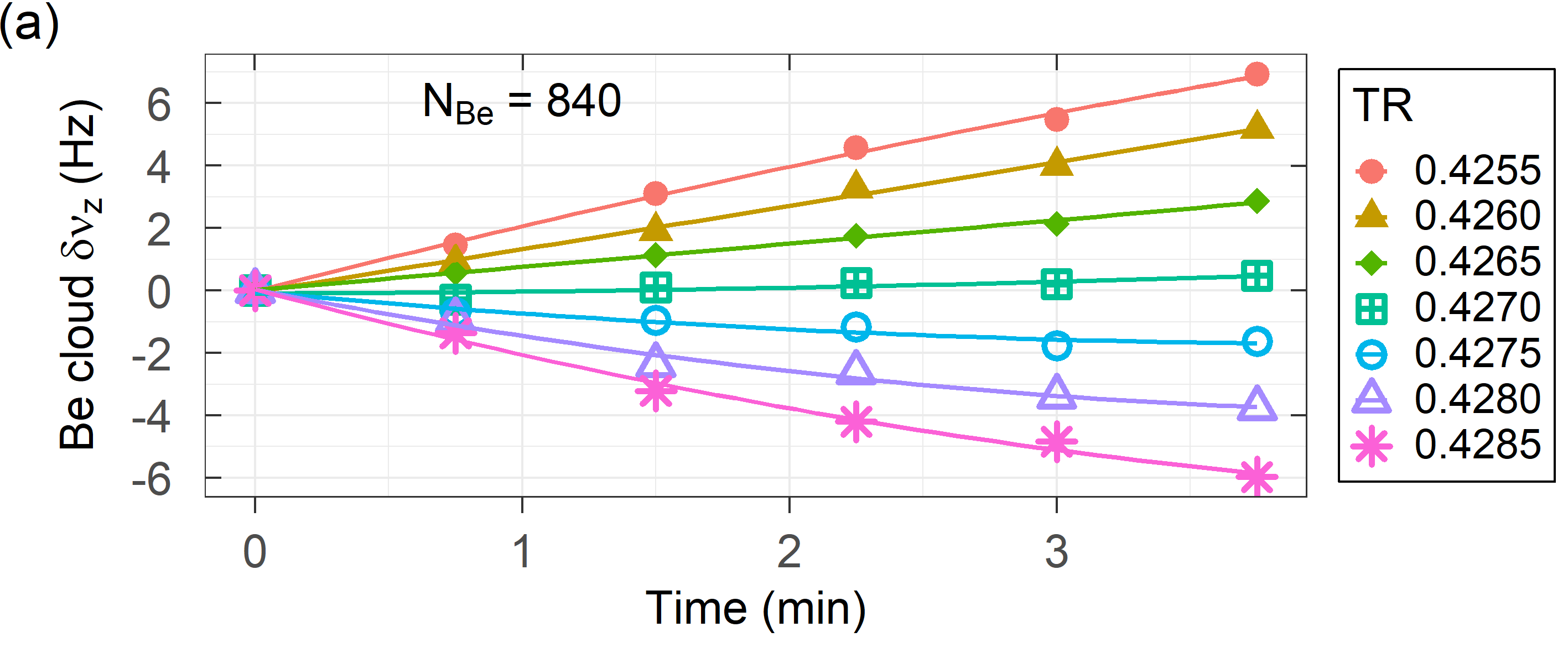} \\
		\includegraphics[width=8.6cm]{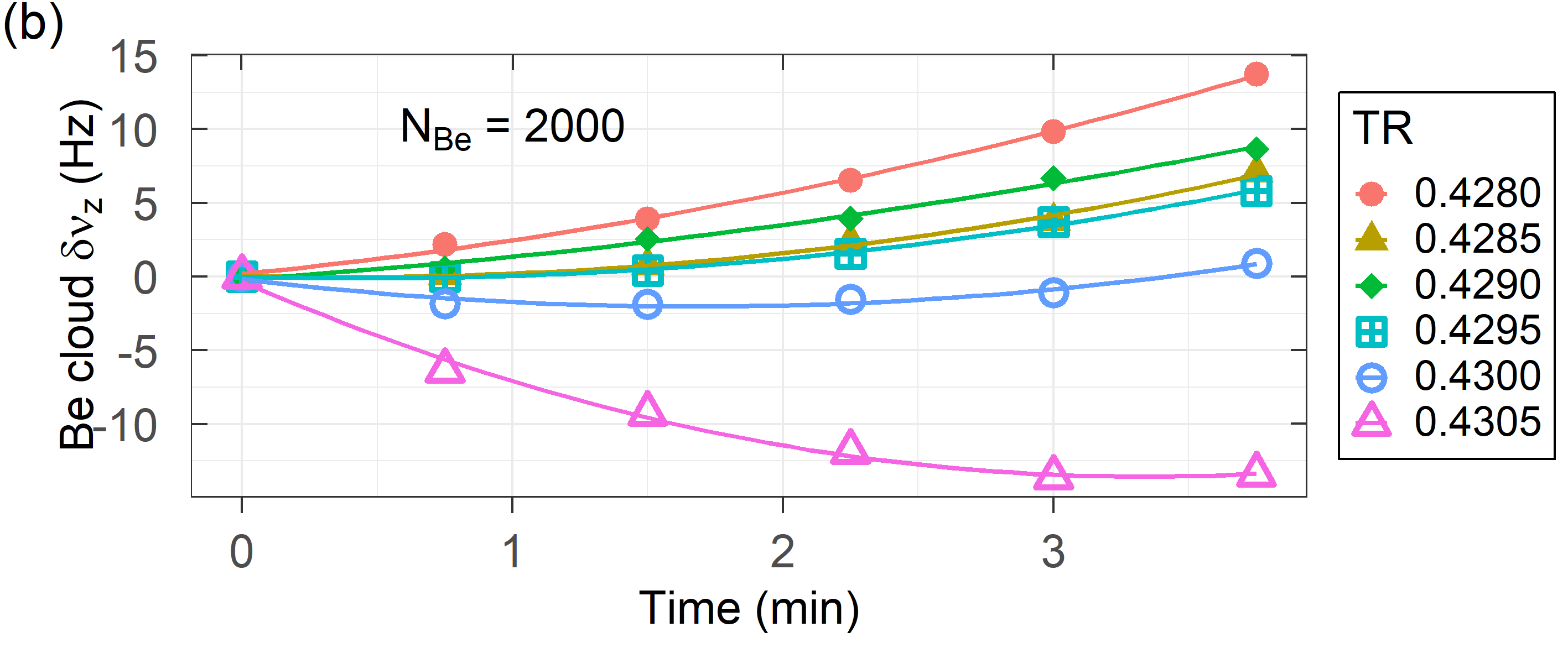} 
	\end{tabular}
	\caption{Axial frequency drift of a cloud of beryllium ions in the LT after applying the magnetron sideband drive at $\nu_\text{RF} = \nu_\text{z,Be} + \nu_\text{-,Be}$. The drift is related to the tuning ratio (TR) of the trap and becomes stronger the larger the cloud is. The TR design value is 0.442 \cite{WieTh}. }
	\label{fig:Be_mag_drift}
\end{figure}

\begin{figure}
	\includegraphics[width=8.6cm]{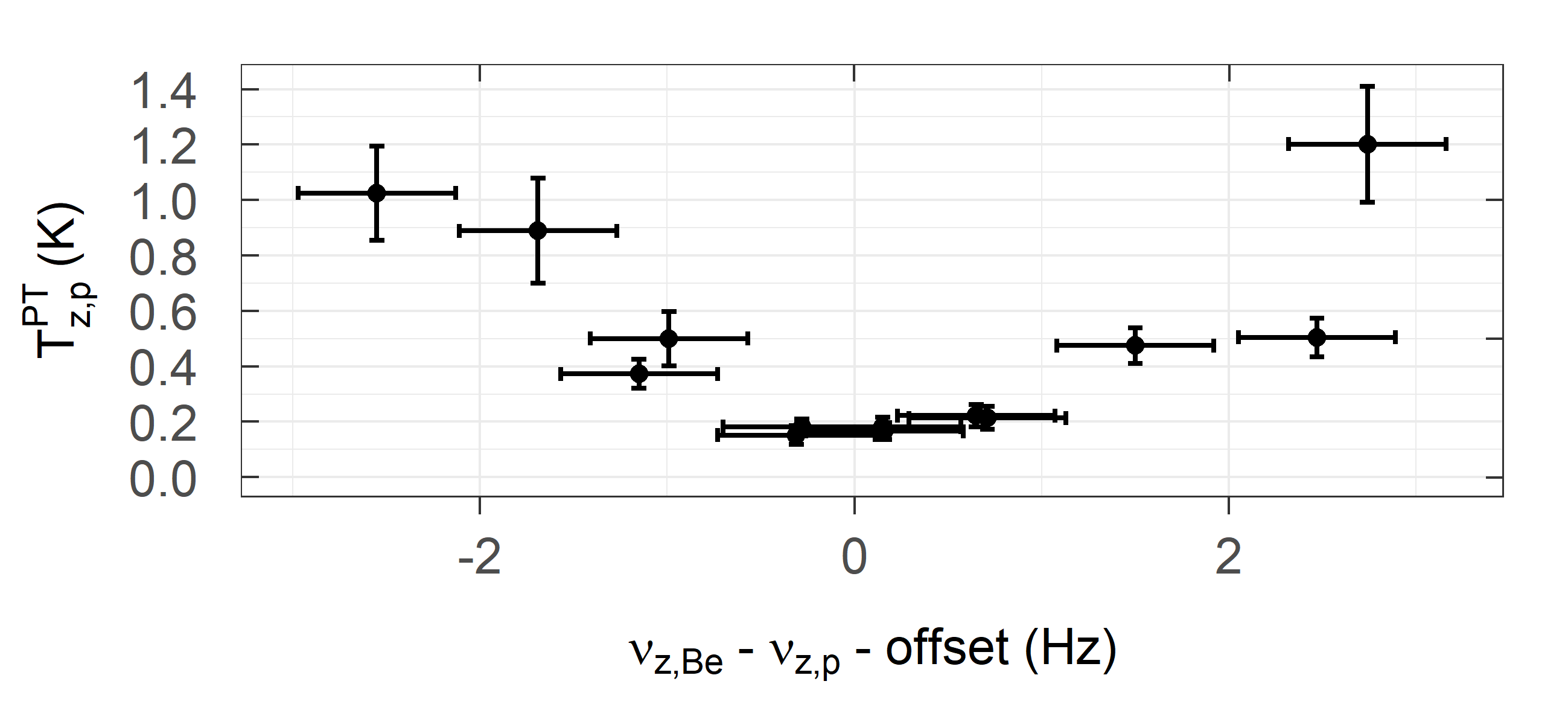}
	\caption{Temperature of the sympathetically cooled proton as a function of relative axial frequency detuning between the proton and beryllium ions. The data was recorded with the 120\,Hz cloud. }
	\label{fig:Tzp_vs_detuning}
\end{figure}

%	x
%	\clearpage
%	\newpage
Laser cooling of atoms and ions is a widely employed method in the field of atomic physics \cite{Coh98, Saf18}. However, only few species offer a suitable optical transition for laser cooling. For most other particles, sympathetic cooling techniques have to be employed.
%	, where the particle of interest is indirectly cooled by coupling it to a laser-coolable species. 
In the established sympathetic cooling schemes, the coupling is realized through direct ion-ion Coulomb interaction, where the charged particles are either trapped in the same potential well \cite{Lar86, Bre19, Mic20} or in two separate ones whose separation distance is a few hundred micrometers only \cite{Har11, Bro11}.  \\
Our group has recently demonstrated the sympathetic laser cooling of a single proton mediated by image currents \cite{Boh21}. Here, the two ion species are placed in two independent Penning traps that are separated by a distance of 5.5\,cm but are connected to the same parallel RLC circuit \cite{Hei90}. Consequently, this technique is especially suited for exotic species such as highly charged ions or charged antimatter \cite{Smo15, Per15}. Furthermore, in comparison to resistive cooling with a  cyclotron resonator that only covers a single fixed particle charge-to-mass ratio $q/m$ \cite{Ulm13, Smo17}, the sympathetic cooling technique is widely tunable and applicable to all relevant $q/m$ simultaneously. 
The lowest axial temperature achieved in Ref.~\cite{Boh21} was $(2.6\pm2.5)$\,K, measured via the temperature-induced axial frequency shift  of about 100\,mHz/K (axial) of the proton in the presence of an anharmonic trapping potential. The temperature resolution of this method is fundamentally limited to about 500\,mK by the typical axial frequency stability of about 50\,mHz.
Furthermore, this temperature measurement method required that the beryllium ion cloud is laser cooled comparably strongly, which is detrimental to the achievable proton temperature \cite{WilTh}. \\
In this Letter, we demonstrate axial temperatures of a single proton in a Penning trap down to about 170\,mK, which constitutes a factor of 15 improvement compared to the previous record. 
To achieve this, we utilize a more precise temperature measurement method based on the axial frequency shift in a quadratic magnetic field inhomogeneity. A new temperature measurement trap (TMT) has been implemented with an axial frequency shift of 470\,Hz/K (axial) \cite{WieTh}. 
\begin{figure*}
	\centering
	\begin{tabular}{cc}
		\rule{0pt}{-1.0ex}%
		\subf{\includegraphics[width=10cm]{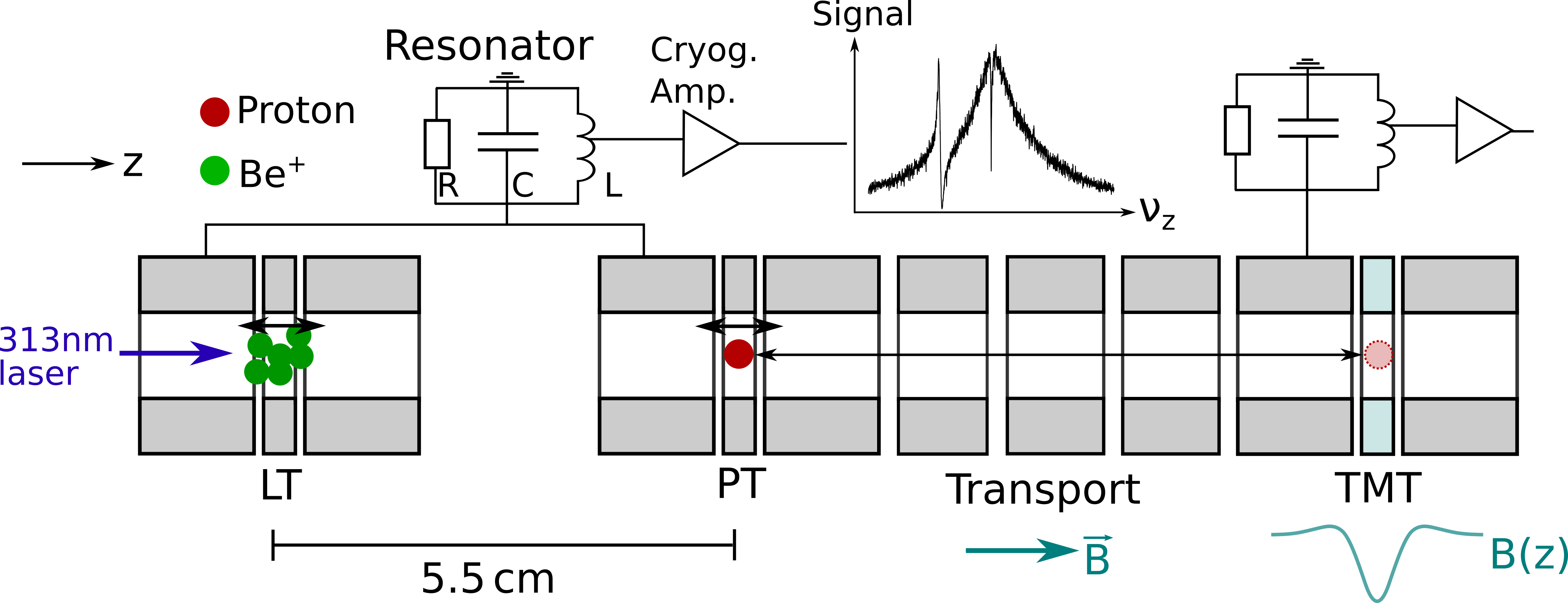}}{ (a) } 
		\subf{\includegraphics[width=6cm]{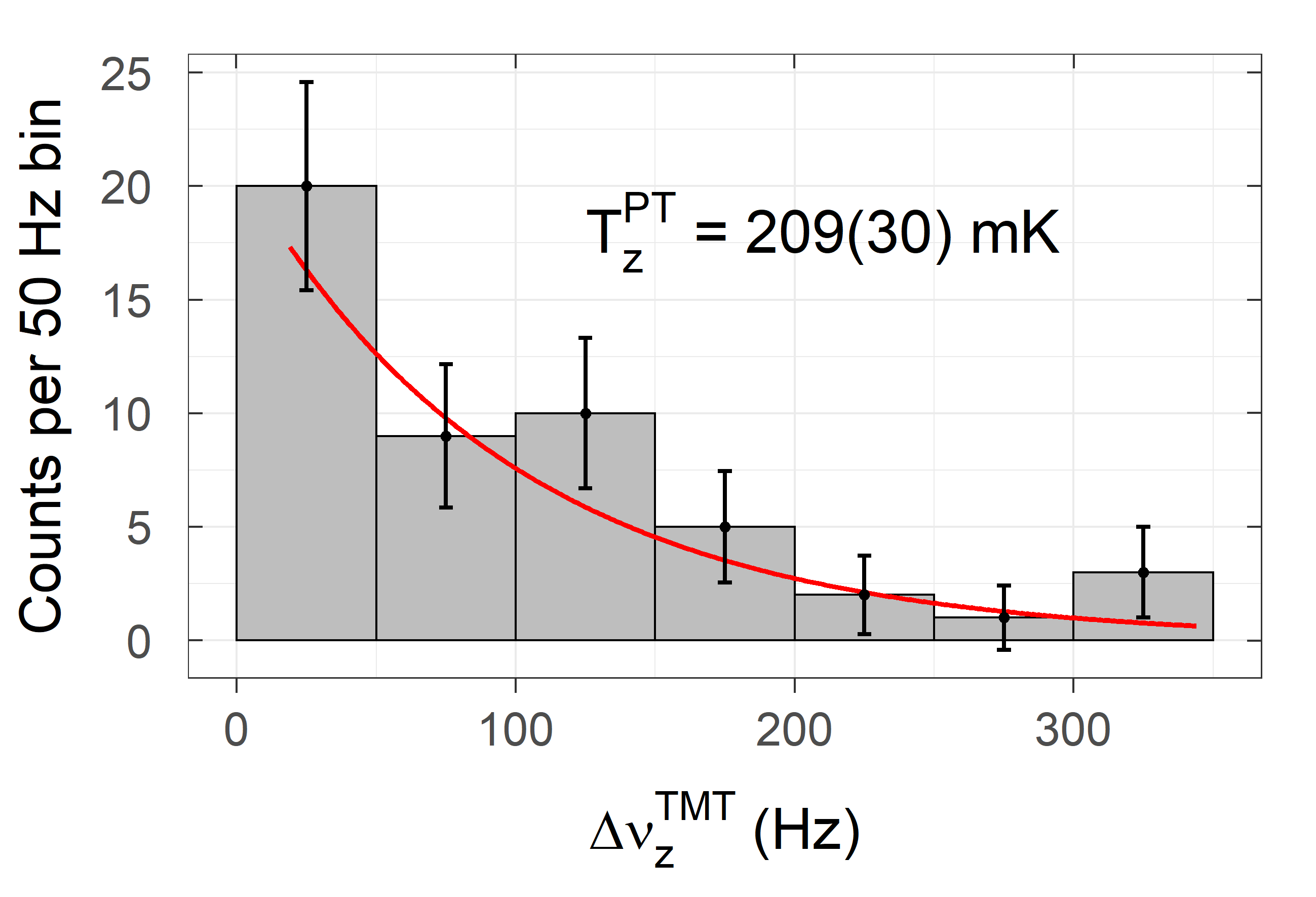}}{ (b) } 
	\end{tabular}
	\caption{(a) Schematic of the experimental setup. A cloud of beryllium ions is trapped in the loading trap (LT) and a single proton in the proton trap (PT). Both traps are connected to a common resonator.
		The proton is transported from the PT into the temperature measurement trap (TMT) for measuring its modified cyclotron energy via a quadratic magnetic field inhomogeneity. (b) Typical Boltzmann distribution of a temperature measurement. The red line corresponds to the Boltzmann distribution with a temperature as determined by the maximum-likelihood method.}
	\label{fig:experimental_setup_and_Boltzmann}
\end{figure*}
The TMT offers a temperature resolution in the mK range as well as a 17-fold faster temperature readout time compared to a preexisting analysis trap (AT) with a larger magnetic field inhomogeneity for spin state readout \cite{Moo13}.
In addition, the understanding of the coupling process has been improved through numerical simulations, which has led to optimized experimental parameters for the cooling process \cite{Wil22, WilTh}.
\\
Our sympathetic cooling experiments are performed in a cryogenic multi-Penning-trap system \cite{WieTh, WilTh} that stores ions by means of a superposition of a homogeneous magnetic field of $B_0 \approx 1.9$\,T and a quadrupolar electric potential, shown schematically in Fig.~\ref{fig:experimental_setup_and_Boltzmann}(a). 
The resulting harmonic motion of a single particle consists of the magnetron and modified cyclotron motion in the radial plane and an axial motion orthogonal to it with frequencies $\nu_-$, $\nu_+$, and $\nu_\z$, respectively \cite{Bro86}.
The axial motion of a particle is detected by tuning its axial frequency to the resonance frequency of a parallel RLC circuit (also called resonator) consisting of a superconducting coil with inductance $L$, a stray capacitance $C$, and an effective parallel resistance $R$ \cite{Win75}.
The resonance is generated by the thermal Johnson-Nyquist noise at a frequency $\nu_0 = 1/(2\pi \sqrt{LC})$ and has a high quality factor $Q=R/(2\pi \nu_0 L)$. For example, $\nu_0^\text{PTLT} = \SI{345250}{Hz}$ and $ Q \approx \SI{10000}{}$ for the resonator that is connected to both the proton trap (PT) and loading trap (LT) in our case \cite{WieTh}. 
%	\red{The particles are detected by tuning their axial frequency to $\nu_0$ by adjusting the trapping potential. }
%	By tuning the axial frequency of a particle to the resonator frequency, the voltage noise of the resonator drives the axial motion of the particle. In turn, the particle induces image currents into the resonator. 
The voltage signal of the resonator is amplified by a cryogenic low-noise amplifier at 4\,K \cite{Nag16} followed by another amplification stage at room temperature.
Recording the fast Fourier transform (FFT) of the resonator voltage signal yields the characteristic spectrum featuring a dip at the ion's axial frequency \cite{Win75}, shown schematically in the inset of Fig.~\ref{fig:experimental_setup_and_Boltzmann}(a).
The coupling of the particles to the resonator is given by their respective dip width $\gamma_\z = \frac{1}{2\pi} \frac{R}{m} \frac{q^2}{D^2} N $, where $N$ is the number of ions of the species and $D$ an effective trap size \cite{Win75}.
The proton dip width in the PT is $2.2$\,Hz per ion and the beryllium dip width in the LT is $0.10$\,Hz per ion. 
The cooling measurements are performed with a single proton trapped in the PT and a cloud of beryllium ions stored in the LT. Both traps are connected to the same resonator, which mediates the energy exchange via image currents between the particles \cite{Boh21}. The proton is sympathetically cooled by tuning the axial frequency of both species to the resonator frequency and laser-cooling the beryllium ions via the 313\,nm $^2$S$_{1/2} \rightarrow ^2$P$_{3/2}$ transition with a natural linewidth of about $\Gamma = 2\pi \times \SI{20}{MHz}$ \cite{And69}. 
\\ In order to measure the temperature of the proton, we employ a two-trap measurement scheme.
Since the resonator constitutes a thermal reservoir,  the axial mode continuously samples a Boltzmann distribution with a correlation time of $1/\gamma_\text{z}$ if $\nu_\z \approx \nu_0$. The corresponding time average is, according to the ergodic theorem, equivalent to the ensemble average which we use as the temperature definition of a single particle. In contrast, the modified cyclotron mode remains at constant energy. 
However, it can be coupled to the axial mode by irradiating a quadrupolar sideband drive at frequency $\nu_\text{RF} = \nu_+ - \nu_\z$ \cite{Cor90}. In this way, an axial energy is imprinted on the modified cyclotron mode. The corresponding temperatures obey the relation $T_+ = T_\z \nu_+/\nu_\z $, where $T_+$ and $T_\z$ are the temperatures of the modified cyclotron mode and the axial mode, respectively. Then, the proton is transported from the PT into a dedicated temperature measurement trap (TMT), where its modified cyclotron energy $E_+$ is measured. The TMT features a ferromagnetic ring electrode made from nickel, which introduces a large quadratic magnetic field inhomogeneity, i.e.~$B_\z^\text{TMT}(z) = B_0^\text{TMT} + B_2^\text{TMT} z^2$. The coefficient $B_2^\text{TMT}$ has been measured to be $B_2^\text{TMT} = \SI{27.8 \pm 0.7}{kT/m^2}$ \cite{WilTh}, which is consistent with the design value \cite{WieTh}. 
To measure $E_+$, we utilize the fact that the $B_2$-coefficient causes an axial frequency shift in the TMT which is proportional to $E_+$ \cite{Bro86, Ket14}, in our case $\delta\nu_\text{z}^\text{TMT}/\delta E_+ = $
5.8\,Hz/(\kkb) for the proton. 
Thus, the axial temperature of the single proton is obtained by repeatedly measuring modified cyclotron energies independently drawn from a Boltzmann distribution. 
An example of such a Boltzmann distribution is given in Fig.~\ref{fig:experimental_setup_and_Boltzmann}(b). \\
In order to convert the distribution of axial frequency measurements in the TMT to an axial temperature in the PT, we employ a maximum-likelihood approach. The maximum-likelihood estimator for the axial temperature in the PT, $T_\text{z}^\text{PT}$, is given by the mean axial frequency shift,
%	\red{Consequently, the axial frequency distribution in the TMT is converted to an axial temperature in the PT, $T_\z^\text{PT}$, by}
\begin{eqnarray}
	T_\z^\text{PT} = \frac{\nu_\z^\text{PT}}{\nu_+^\text{PT} } \frac{B_0^\text{PT}}{B_2^\text{TMT}} \frac{4\pi^2}{\kb} m \, \nu_\text{z,0}^\text{TMT}  \braket{ \nu_\z^\text{i,TMT} - \nu_\text{z,0}^\text{TMT} }.
	\label{eq:Tz_PT}
\end{eqnarray}
Here, $\nu_\text{z,0}^\text{TMT} \approx \SI{550875}{Hz}$ is the unshifted axial frequency in the TMT at $E_+ = 0$ and $\nu_\z^\text{i,TMT}$ are the individual axial frequency measurements. $B_0^\text{PT} = 1.899$\,T is the magnetic field in the PT and $\nu_+^\text{PT} \approx \SI{28.9}{MHz} $ and $\nu_\z^\text{PT} = \SI{345250}{Hz}$ are the modified cyclotron and axial frequencies of the proton in the PT, respectively, and $\kb$ is Boltzmann's constant.
This formula incorporates not only the axial frequency shift due to non-zero $E_+$, but also the temperature relation due to sideband coupling and the relative modified cyclotron energy change during transport into a different magnetic field, in our case by the magnetic field ratio of the traps, $B_0^\text{PT}/B_0^\text{TMT}$. 
\\
In support of the experimental effort, we performed first-principles simulations
of the experimental setting that is shown in Fig.~\ref{fig:experimental_setup_and_Boltzmann}(a) \cite{Wil22, WilTh}.
We found that the on-resonance cooling scheme employed in this work, where the proton and beryllium ion cloud are tuned to the resonator frequency, can be understood by the formation of a symmetric and antisymmetric normal mode of the axial motion of the coupled proton-beryllium system. 
The antisymmetric mode decouples from the resonator, so that the laser cools it close to the Doppler limit.
In contrast, the symmetric mode couples and thermalizes to the resonator and the relative proton and $^9$Be$^+$ component are given by their respective dip widths. Since image-current coupling relies on the motion of the particles, for an optimal proton temperature the damping rate of the cooling laser  $\gamma_\text{L}$ must be sufficiently weak in order to not decouple the beryllium ions from the resonator, i.e.~$\gamma_\text{L} \ll \gamma_\text{z,Be}$, where $\gamma_\text{z,Be}$ denotes the dip width of the beryllium cloud.
Then, the axial temperature of a proton with dip width $\gamma_\text{z,p}$ is given by \cite{Wil22, WilTh}
\begin{eqnarray}
	T_\text{z,p} = \frac{1}{1 + \frac{\gamma_\text{z,Be}}{\gamma_\text{z,p}}} T_\text{res},
	\label{eq:Tzp_prediction}
\end{eqnarray}
where $T_\text{res}$ is the effective resonator temperature. 
Thus, besides $T_\text{res}$, the ratio of the proton and beryllium dip widths determines the final proton temperature. 
In order to compare theory with experiment,	$T_\text{res}$ is measured as the first preparatory step to $T_\text{res} = (8.6 \pm 0.8)$\,K.
The cryogenic amplifier is turned off for this measurement as well as for the sympathetic cooling measurements, since otherwise the amplifier's input noise gives rise to a slightly higher effective axial temperature. \\
It is crucial to minimize the axial frequency difference of the two species for efficient sympathetic cooling. Hence, in the following their individual frequency stabilites are examined.
The axial frequency stability of the single proton in the PT is $\sigma(\nu_\text{z,p}) \approx 40$\,mHz for 60\,s averaging time, which is negligibly small for the sympathetic cooling.
In contrast, the axial frequency stability of the beryllium cloud is adversely affected by the radial cloud expansion due to the Coulomb interaction \cite{Bol93, Wei94}. 
In order to prevent an uncontrolled radial expansion, the radial modes of a large beryllium cloud must be regularly cooled. 
However, in our setup the laser is applied nearly parallel to the trap axis so that the cooling of radial modes \cite{Ita82} is not efficient. Since large laser powers are not feasible because the cooling scheme requires low photon scattering rates, a magnetron sideband drive at $\nu_\text{RF} = \nu_\text{z,Be} + \nu_\text{-,Be}$ \cite{Cor90} is applied in addition to the cooling laser. 
We observe axial frequency drifts of the beryllium ion clouds after turning off the magnetron sideband drive, as shown exemplarily in Fig.~\ref{fig:Be_mag_drift} for two ion clouds of 840 ions and 2000 ions \cite{WieTh, WilTh}.
These drifts are related to the harmonicity of the electrical trap potential, which is adjustable via the tuning ratio (TR) of the trap as defined in Ref.~\cite{Bro86}.
We attribute these frequency drifts to the change in the aspect ratio of the beryllium ion cloud while the cloud expands due to Coulomb repulsion in an anharmonic trapping potential \cite{Wei94}. 
The accumulated frequency shift can be reset by applying the magnetron sideband drive again.
Figure\,\ref{fig:Be_mag_drift}(a) demonstrates that a TR optimization allows to control the frequency drifts of clouds of $\lesssim$ 1000 ions to $<1$\,Hz on the relevant time scale, which is sufficiently stable for the sympathetic cooling process. 
In contrast, for larger clouds the drifts are not only stronger, they also exhibit a quadratic component, as shown in Fig.~\ref{fig:Be_mag_drift}(b). 	For large clouds above 1000 ions, this requires us to introduce small frequency offsets and apply the coupling procedure during the times when the change in detuning between the proton and beryllium ions is minimal. 
\\ With a precisely adjusted TR, we were able to sufficiently stabilize beryllium ion clouds with dip widths of up to 120\,Hz or 1200 ions. Assuming the proton is initially trapped in the PT, one sympathetic cooling cycle consists of the following steps: First, in order to start the cooling process, the axial frequencies of the proton and the beryllium ion cloud are matched in between two iterations of applying the magnetron sideband drive to the beryllium ion cloud.
The cooling laser is kept continuously on with a weak damping rate during the whole cycle.  During the sympathetic cooling,  the modified cyclotron sideband frequency of the proton is continuously applied at $\nu_\text{RF} = \nu_\text{+,p} - \nu_\text{z,p}$ with about 1\,Hz Rabi frequency. 
The total cooling time is set to 90\,s. The cooling time constant has not been explicitly measured, however, it was verified that after 90\,s the resulting temperatures have converged and two subsequent values of $E_+$ are uncorrelated. Afterwards, the proton is transported into the TMT to measure $E_+$. Finally, the proton is transported back into the PT and the next cooling cycle is carried out.
A single temperature measurement consists of $>20$ individual cooling cycles.
A more detailed description of the individual steps can be found in Ref.~\cite{WilTh}. \\
\begin{figure}
	\centering
	\begin{tabular}{cccc}
		\rule{0pt}{-1.0ex}%
		\includegraphics[width=4.15cm]{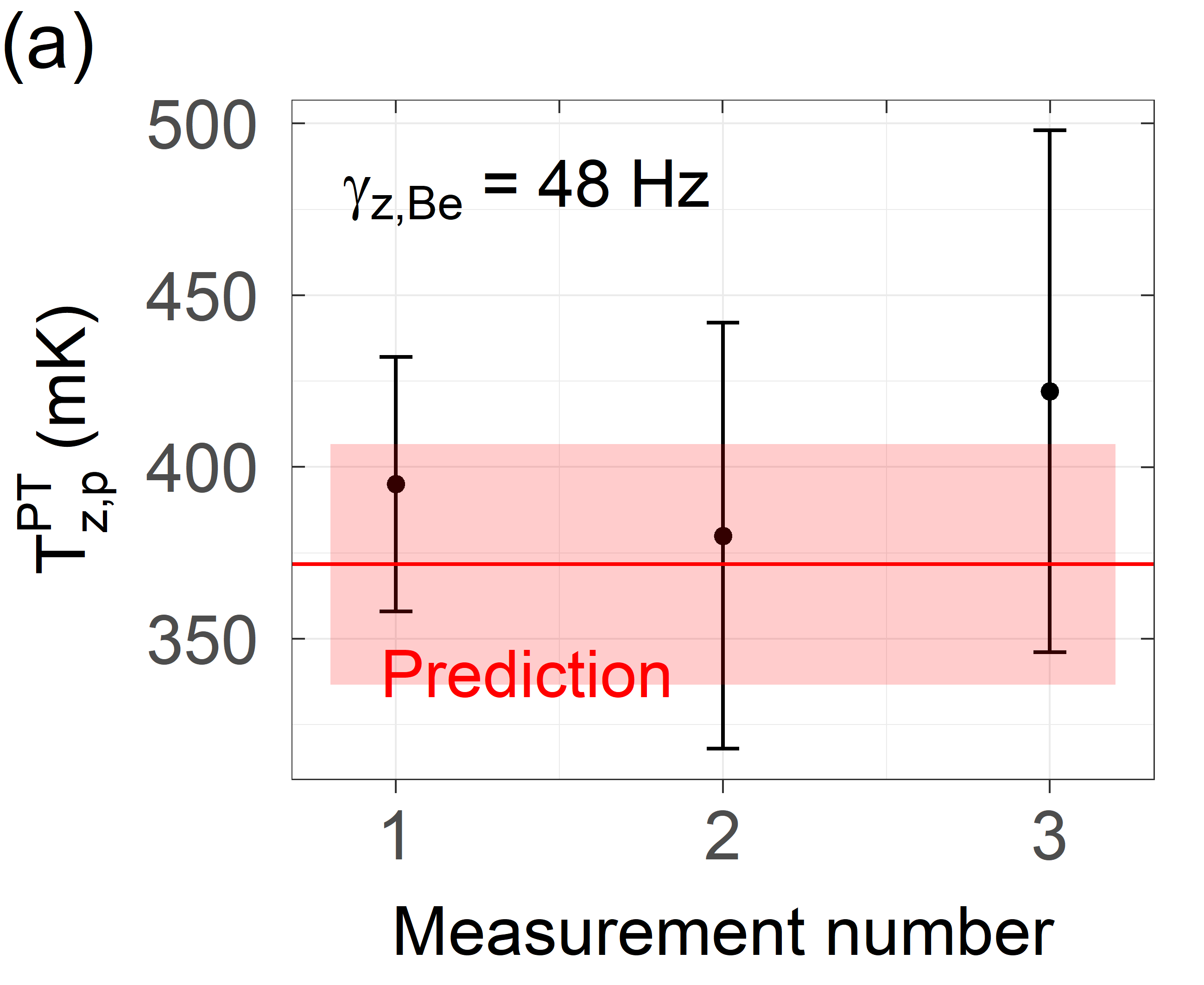} \hspace{0.3cm}
		\includegraphics[width=4.15cm]{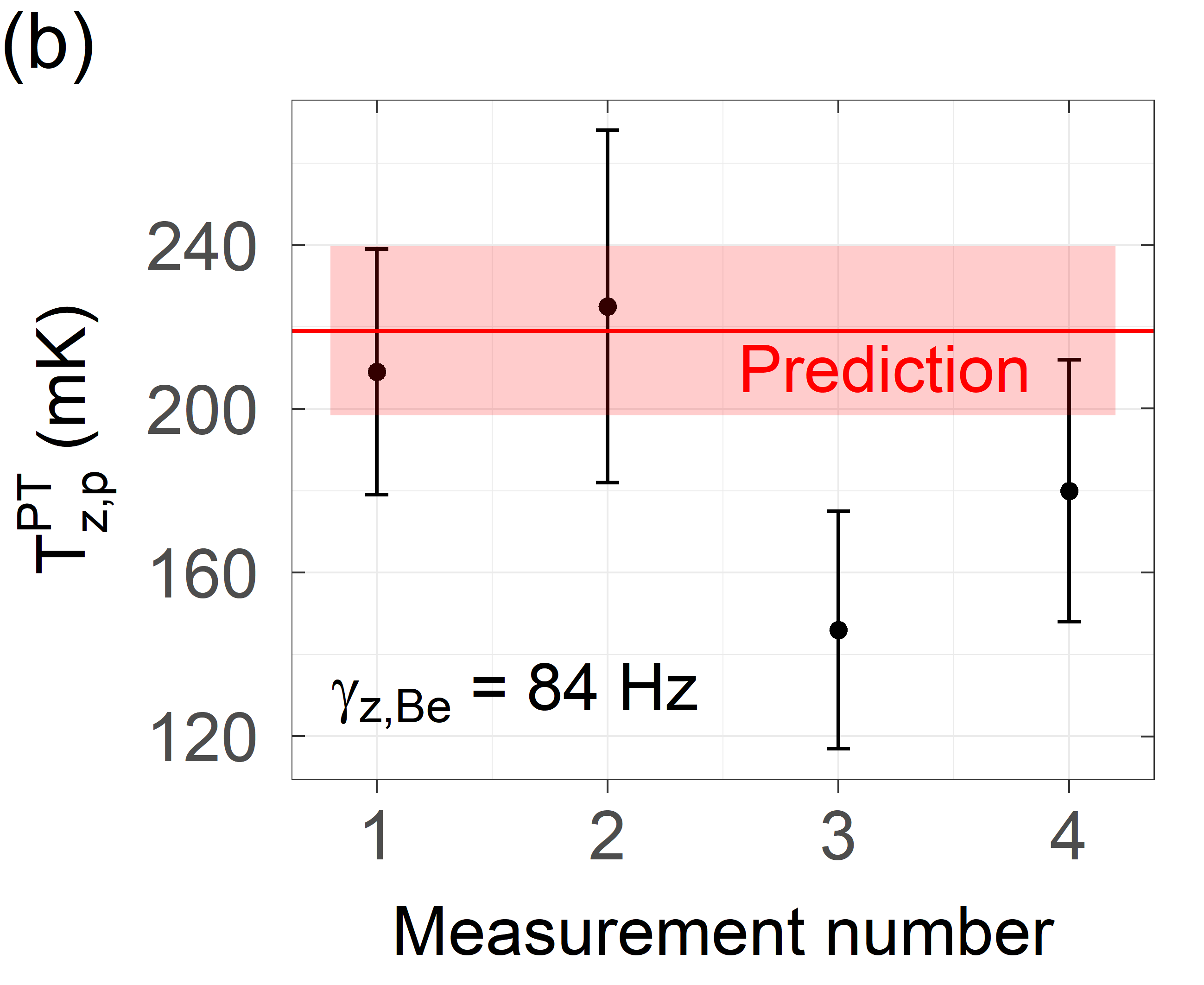} \\
		\includegraphics[width=4.15cm]{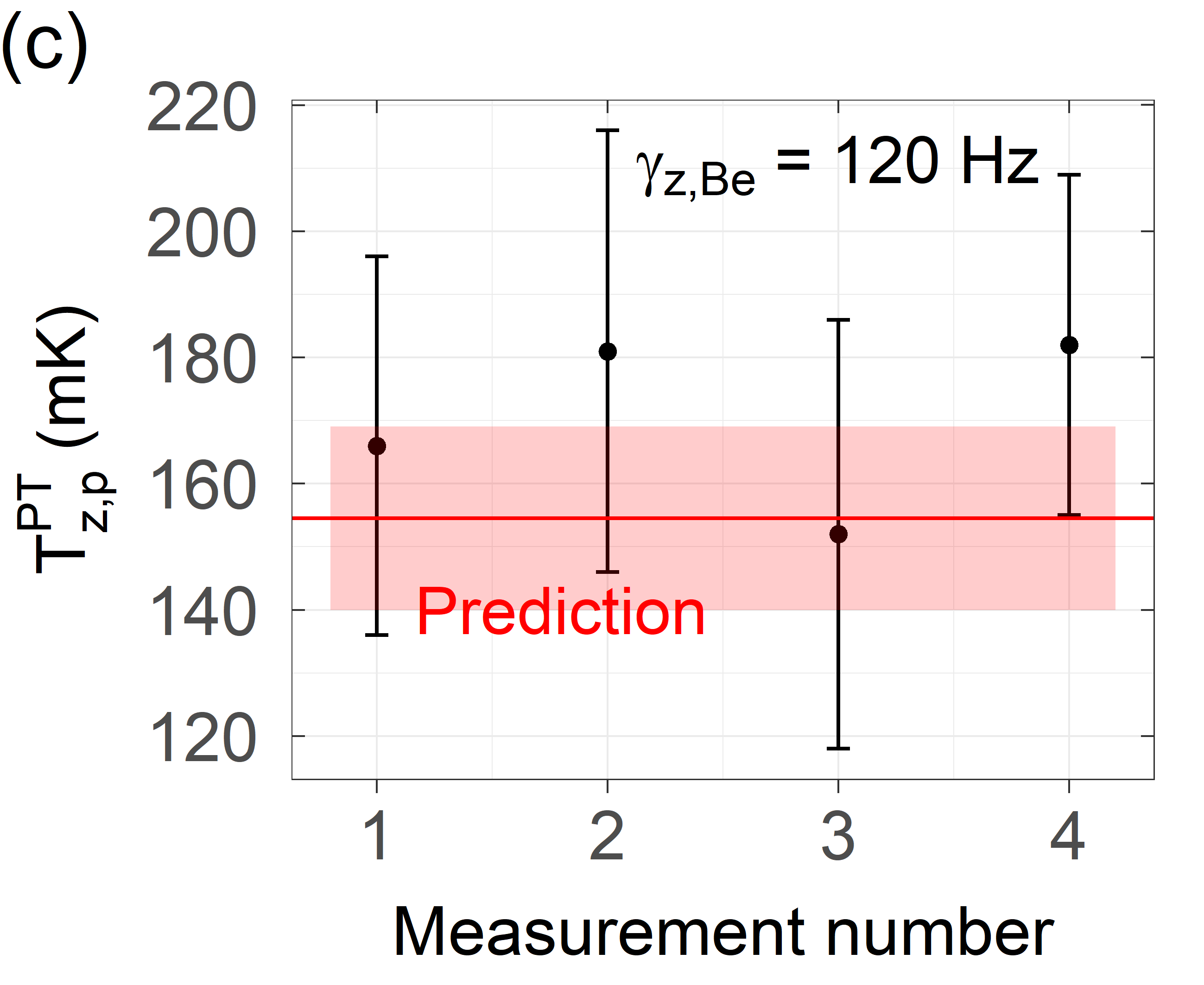} \hspace{0.3cm}
		\includegraphics[width=4.15cm]{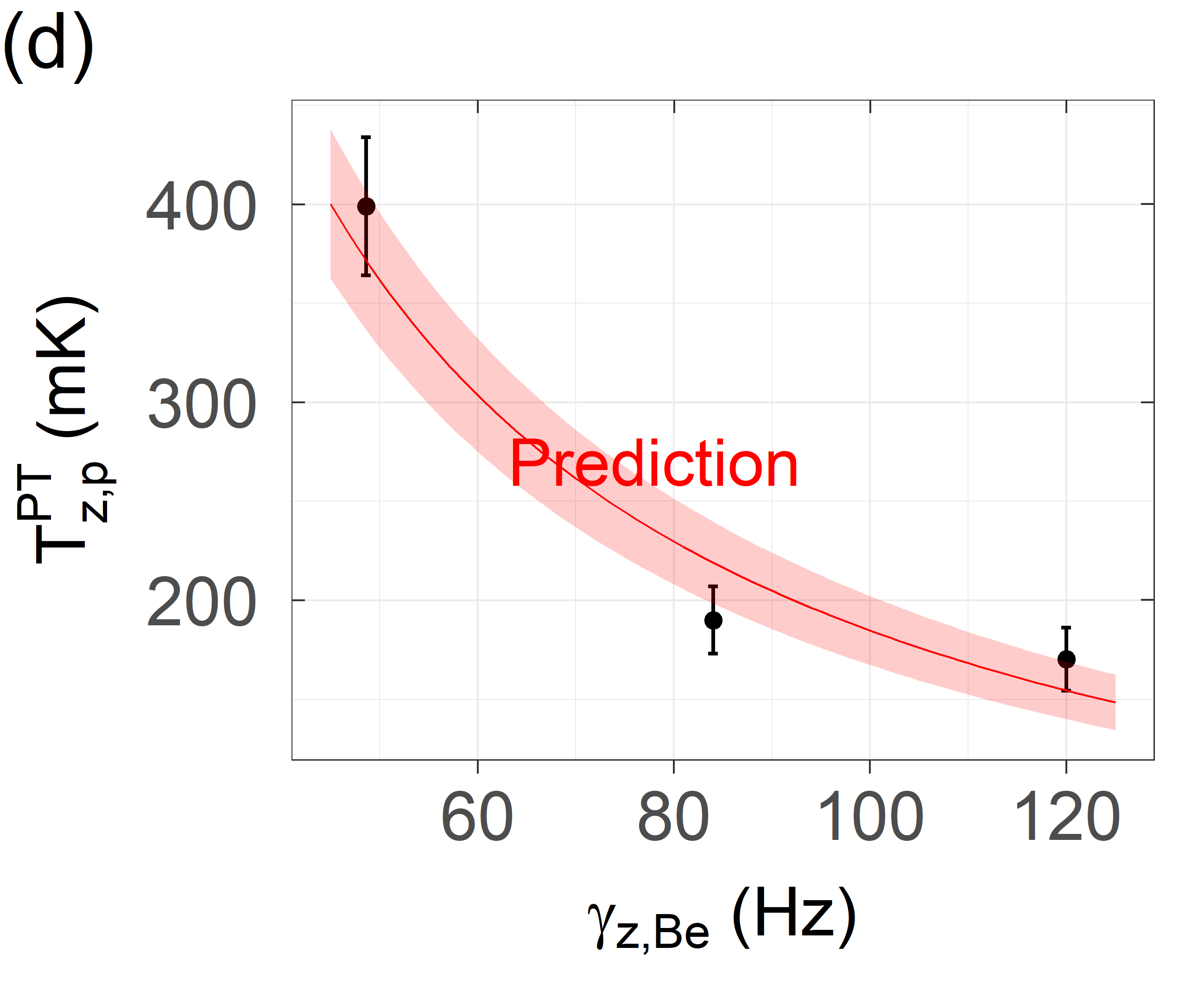} 
		%			\subf{\includegraphics[width=3.5cm]{cooling_measurements_84Hz_22_12.png}}{ (b) } \\
		%			\subf{\includegraphics[width=8.6cm]{cooling_measurements_120Hz_22_12.png}}{ (c) }  \\
		%			\subf{\includegraphics[width=8.6cm]{cooling_measurements_Tz_vs_Bedipwidth_22_12.png}}{ (d) } 
	\end{tabular}
	\caption{Axial temperature measurements of a sympathetically cooled proton. Three differently sized beryllium ion clouds were employed. The red line is the prediction by Eq.~(\ref{eq:Tzp_prediction}) and the shaded red area the associated uncertainty. In (d) the individual measurements for each cloud size are combined and plotted as a function of the beryllium ion cloud dip width $\gamma_\text{z,Be}$.  }
	\label{fig:cooling_measurements}
\end{figure}
This cooling process has been conducted for three differently sized beryllium ion clouds, namely with dip widths of 48\,Hz, 84\,Hz, and 120\,Hz, corresponding to 480, 840, and 1200 ions. 
For the 48\,Hz cloud, the cooling laser was set to 84\,MHz red detuning and its power was adjusted to 5--\SI{15}{\micro W}, which corresponds to 1.5--4.5\% of the saturation power of the cooling transition. The saturation power of $P_\text{sat} = \SI{340}{\micro W}$ was measured by in-trap detection of fluorescence photons \cite{Wie23arxiv, WieTh}.
For the 84\,Hz and 120\,Hz clouds, the laser power and red detuning were increased to \SI{100}{\micro W} and 200\,MHz, respectively.
In contrast to our previous work \cite{Boh21}, both parameter sets fulfill the condition that the beryllium ion cloud is damped only weakly and that no loss of the SNR of the beryllium dip occurs.
Several independent proton temperature measurements have been conducted for each beryllium cloud with results shown in Fig.~\ref{fig:cooling_measurements}, where the uncertainty of each temperature measurement is dominated by the statistical uncertainty. 
The horizontal red line is the temperature predicted by Eq.~(\ref{eq:Tzp_prediction}) and the shaded red area is the corresponding uncertainty, which is dominated by the statistical uncertainty of the measurement of $T_\text{res}$.
In Fig.~\ref{fig:cooling_measurements}(d) the individual measurements are combined and plotted as a function of the dip width and thus particle number of the beryllium ion cloud.
In general, we observe excellent agreement between the theoretical prediction and the experimental data. The lowest reproducibly measured temperature of about 170\,mK constitutes a 15-fold improvement compared to the previous record measurement \cite{Boh21}. We emphasize that not only the temperatures agree, but also the method to reach them: The prediction by the simulations that a weak laser damping rate is required \cite{Wil22} has been confirmed as well. \\
Moreover, our results demonstrate the capability of the two-trap temperature measurement technique. By separating the temperature determination from the cooling process a broad range of temperatures between 5\,mK and 10\,K can be measured and resolved.
In this regard we also measured a negligible heating rate of $\lesssim 2$\,mK per cycle, where we repeated the cooling sequence but detuned the modified cyclotron sideband frequency for the proton by 10\,kHz \cite{WilTh}. 
\\
Further, we study the effect of a relative axial frequency detuning between the proton and the beryllium ion cloud. 
To this end, the 120\,Hz cloud was tuned to slightly different axial frequencies than the proton. All other steps remain the same as before. The resulting proton temperatures are shown in Fig.~\ref{fig:Tzp_vs_detuning}.
Notably, these measurements were also recorded with the cryogenic amplifier turned off, so that the relative frequency detuning is estimated based on the initial center frequency as well as the TR-related frequency drift. This gives rise to a small offset of about 2\,Hz. As a result, the detunings should only be considered a coarse estimate. Nevertheless, the existence of a cooling resonance is evident and the minimum corresponds to the prediction by Eq.~(\ref{eq:Tzp_prediction}).
\\	To achieve lower proton temperatures, an alternative to increasing the number of beryllium ions is to reduce the resonator temperature. Thus, we have also conducted temperature measurements where the effective resonator temperature was reduced to $T_\text{res}^\text{fb} = (2.6 \pm 0.3)$\,K by the application of negative feedback \cite{Dur03}. However, here the proton temperature was limited to $\geq 300$\,mK independent of the number of beryllium ions \cite{WilTh}, which is a strong indication for a feedback-induced noise limitation.
\\ In summary, we have demonstrated image-current mediated sympathetic cooling of a single proton to axial temperatures down to about 170\,mK, an improvement by a factor of 15 compared to the previous record \cite{Boh21}.
As such, this work constitutes a crucial milestone towards the next generation of high-precision Penning trap  measurements with particles that require sympathetic cooling with separate trapping regions. \\ 
Several routes towards lower proton temperatures are conceivable: 
An optimized beryllium ion trap for which the dip width per Be$^+$ is maximized and the anharmonicity-related frequency drifts are minimized would directly enable lower proton temperatures \cite{BohTh, WieTh, WilTh}. Alternatively, with an independent cooling laser in radial direction it would be possible to control and stabilize significantly larger beryllium ion clouds via laser cooling only. 
Then, the magnetron sideband coupling with its associated frequency drifts would become obsolete. 
Another option would be to confine large ion clouds radially with a rotating wall potential \cite{Bha12}, which could enable the use of significantly larger beryllium ion clouds as well. 
Besides, the simulation studies \cite{Wil22} as well as independent work in Ref.~\cite{Tu21} predict that a cooling scheme with several kHz particle-resonator detuning and pulsed laser cooling can achieve 10\,mK axial particle temperatures with about 100 beryllium ions only. 
The excellent agreement between experiment and simulation in this work further corroborates the fundamental feasibility of these cooling methods.
\\ Regardless which cooling scheme turns out to be the most suitable one for further temperature reduction, the fact that all of them rely on image-current coupling makes them in principle applicable to any trapped charged particle and experimental systems beyond Penning traps \cite{An22}. Consequently, these cooling methods are of special interest for charge-, parity-, and time (CPT) symmetry tests with protons and antiprotons \cite{Smo15}, magnetic moment measurements of light nuclei \cite{Moo18}, as well as high-precision mass measurements \cite{Rep12} and tests of quantum electrodynamics with highly charged ions \cite{Tu21, Stu19, Egl19} in Penning traps. In particular, once even lower temperatures of about 10\,mK (axial) can be reached \cite{Cor23}, the sympathetic cooling will significantly boost the sampling rate and spin state detection fidelity \cite{Smo17_PhysLettB, Boh17} of future $g$-factor measurements on protons \cite{Sch17}, antiprotons \cite{Smo17} and other nuclear moments, as well as reduce the dominant systematic uncertainties in mass measurements with the highest precision \cite{Bor22}.

\section{Acknowledgements}
This study comprises parts of the PhD  thesis work of C. Will and M. Wiesinger. 
We acknowledge financial support from 
the Max-Planck-Society, 
the RIKEN Chief Scientist Program, 
the RIKEN Pioneering Project Funding, 
the RIKEN JRA Program, 
the Royal Society,
the Helmholtz-Gemeinschaft, 
the DFG through SFB 1227 ‘DQ-mat’, 
the cluster of excellence QuantumFrontiers,
the CERN Gentner programme, 
the Max-Planck IMPRS-PTFS, 
the European ResearchCouncil (ERC) under the European Union’s Horizon 2020 research and innovation programme (Grant agreement Nos. 832848 -- FunI, 721559 -- AVA, 852818 -- STEP), 
and the Max-Planck–RIKEN–PTB Center for Time, Constants and Fundamental Symmetries.

	\bibliographystyle{apsrev4-2} % Tell bibtex which bibliography style to use
	\bibliography{references}% Produces the bibliography via BibTeX.
	
\end{document}